\documentclass[aps,prd,eqsecnum,showpacs,amsmath]{revtex4}
\usepackage[dvips]{color,graphicx}
\usepackage{amsfonts}
\usepackage{amssymb}
\usepackage{latexsym}

\newcommand{\dalm}{\kern1pt\vbox{\hrule height 0.9pt\hbox{\vrule width
0.9pt\hskip 2.5pt\vbox{\vskip 5.5pt}\hskip 3pt\vrule width 0.3pt}\hrule height
0.3pt}\kern1pt}

\newtheorem{The}{Theorem}
\newtheorem{lm}{Lemma}
\newtheorem{dn}{Definition}
\newtheorem{Coro}{Corollary}

\begin{document}


\title{
Final fate of spherically symmetric gravitational collapse of a dust cloud in Einstein-Gauss-Bonnet gravity}
\author{
Hideki Maeda
\footnote{Electronic address: hideki@gravity.phys.waseda.ac.jp}
}

\address{ 
Advanced Research Institute for Science and Engineering,
Waseda University, Okubo 3-4-1, Shinjuku, Tokyo 169-8555, Japan
}

\date{\today}

\begin{abstract}                
We give a model of the higher-dimensional spherically symmetric gravitational collapse of a dust cloud including the perturbative effects of quantum gravity.
The $n (\geq 5)$-dimensional action with the Gauss-Bonnet term for gravity is considered and a simple formulation of the basic equations is given for the spacetime $M \approx M^2 \times K^{n-2}$ with a perfect fluid and a cosmological constant.
This is a generalization of the Misner-Sharp formalism of the four-dimensional spherically symmetric spacetime with a perfect fluid in general relativity.
The whole picture and the final fate of the gravitational collapse of a dust cloud differ greatly between the cases with $n=5$ and $n \ge 6$.
There are two families of solutions, which we call plus-branch and the minus-branch solutions. 
A plus-branch solution can be attached to the outside vacuum region which is asymptotically anti-de Sitter in spite of the absence of a cosmological constant.
Bounce inevitably occurs in the plus-branch solution for $n \ge 6$, and consequently singularities cannot be formed. 
Since there is no trapped surface in the plus-branch solution, the singularity formed in the case of $n=5$ must be naked. 
On the other hand, a minus-branch solution can be attached to the outside asymptotically flat vacuum region.
We show that naked singularities are massless for $n \ge 6$, while massive naked singularities are possible for $n=5$.
In the homogeneous collapse represented by the flat Friedmann-Robertson-Walker solution, the singularity formed is spacelike for $n \ge 6$, while it is ingoing-null for $n=5$.
In the inhomogeneous collapse with smooth initial data, the strong cosmic censorship hypothesis holds for $n \ge 10$ and for $n=9$ depending on the parameters in the initial data, while a naked singularity is always formed for $5 \le n \le 8$.
These naked singularities can be globally naked when the initial surface radius of the dust cloud is fine-tuned, and then the weak cosmic censorship hypothesis is violated.
\end{abstract}


\maketitle

\section{Introduction}
Einstein's general theory of relativity has successfully passed many observational tests and is now a central paradigm in gravitation physics.
General relativity explains such gravitational phenomena as the perihelion shift of Mercury's orbit, gravitational lensing, redshift in the light spectrum from extragalactic objects, and so on.
One of the most intriguing predictions of the theory is the existence of a spacetime region from which nothing can escape, i.e., a black hole.

It has been considered that black holes are formed from the gravitational collapse in the last stage of heavy stars' life or in high-density regions of the density perturbations in the early universe.
The first analytic model of black-hole formation in general relativity was obtained by Oppenheimer and Snyder in 1939, which represents the spherically symmetric gravitational collapse of a homogeneous dust cloud in asymptotically flat vacuum spacetime~\cite{os1939}.
In this spacetime, the singularity formed is spacelike and hidden inside the black-hole event horizon, so that it is not visible to any observer.
However, it was shown later that this is not a typical model and the singularities formed in generic collapse are {\em naked}, i.e., observable~\cite{christodoulou1984,newman1986,ns-dust,nm2002}.

In general relativity, it was proven that spacetime singularities inevitably appear in general situations and under physical energy conditions~\cite{he1973}. 
Gravitational collapse is one of the presumable scenarios in which singularities are formed.
Where a naked singularity exists, the spacetime is not globally hyperbolic, so that the future predictability of the spacetime breaks down.
In this context, Penrose proposed the {\it cosmic censorship hypothesis} (CCH), which prohibits the formation of naked singularities in gravitational collapse of physically reasonable matters with generic regular initial data~\cite{penrose1969,penrose1979}.
The weak version of CCH prohibits only the formation of globally naked singularities, i.e., those which can be seen by an observer at infinity.
If the weak CCH is correct, singularities formed in generic gravitational collapse are hidden inside black holes, and the future predictability of the spacetime outside the black-hole event horizon is guaranteed.
On the other hand, the strong version of CCH prohibits the formation of locally naked singularities also, which can be seen by some local observer.
The strong CCH asserts the future predictability of the whole spacetime, i.e., global hyperbolicity of the spacetime.

The CCH is one of the most attractive and important unsolved problems in gravitation physics.
Validity of the CCH is assumed in the many strong theorems such as the black-hole uniqueness theorem or the positive energy theorem in asymptotically flat spacetime.
At present, however, the general proof of the CCH is far from complete.
On the contrary, there are many counterexample ``candidates'' in general relativity. (See~\cite{harada2004} for a review.) 

The formation of a singularity means that a spacetime region with infinitely high curvature can be realized in the very final stage of gravitational collapse.
It is naturally considered that quantum effects of gravity cannot be neglected in such regions, so that the classical theory of gravity cannot be applied there.
Therefore, naked singularities give us a chance to observe the quantum effects of gravity.
From this point of view, Harada and Nakao proposed a concept named the {\it spacetime border}, which is the spacetime region where classical theories of gravity cannot be applied~\cite{hn2004}.
The spacetime border is an effective ``singularity'' in classical theory, and then the CCH can be naturally modified to more a practical version, which prohibits the formation of naked spacetime borders. 
If the modified CCH is true, spacetime regions where quantum effects of gravity dominate are never observed.
On the other hand, if it is violated, there is a possibility in principle for us to observe such regions and obtain information useful to the construction of the quantum theory of gravity, which is still incomplete.
From this point of view, studies of the final fate of gravitational collapse are quite important.

Up to now, many quantum theories of gravity have been proposed.
Among them, superstring/M-theory is the most promising candidate and has been intensively investigated, which predicts higher-dimensional spacetime (more than four dimensions).
In this theory, when the curvature radius of the central high-density region in gravitational collapse is comparable with the compactification radius of extra dimensions, the effects of extra dimensions will be important.
Such regions can be modeled effectively by higher-dimensional gravitational collapse.

A recent attractive proposal for a new picture of our universe, which is called the braneworld universe~\cite{large,Randall,DGP}, is based on superstring/M-theory~\cite{Lukas}. 
In the braneworld universe, we live on a four-dimensional timelike hypersurface embedded in the higher-dimensional bulk spacetime. 
Because the fundamental scale could be around the TeV scale in this scenario, the braneworld suggests that the creation of tiny black holes in the upcoming high-energy collider is possible~\cite{Bhformation}. 
From this point of view, the effects of superstring/M-theory on black holes or gravitational collapse should be investigated.

However, the non-perturbative aspects of superstring/M-theory are not understood completely so far, although the progress in recent years has been remarkable.
Given the present circumstances, taking their effects perturbatively into classical gravity is one possible approach to studying the quantum effects of gravity.
The Gauss-Bonnet term in the Lagrangian is the higher curvature correction to general relativity and naturally arises as the next leading order of the $\alpha'$-expansion of heterotic superstring theory, where $\alpha'$ is the inverse string tension~\cite{Gross}.
Such a theory is called the Einstein-Gauss-Bonnet gravity.

In a previous paper, the author presented a model of the $n(\ge 5)$-dimensional spherically symmetric gravitational collapse of a null dust fluid in Einstein-Gauss-Bonnet gravity~\cite{maeda2005}.
It was shown that the spacetime structure of the gravitational collapse differs greatly between $n=5$ and $n \ge 6$.
In five dimensions, massive timelike naked singularities can be formed, which never appear in the general relativistic case, while massless ingoing-null naked singularities are formed in the $n(\ge 6)$-dimensional case.

In this paper, we consider the $n(\ge 5)$-dimensional spherically symmetric gravitational collapse of a dust fluid with smooth initial data in Einstein-Gauss-Bonnet gravity.
In general relativity, the same system has been analyzed by many researchers both for $n=4$~\cite{christodoulou1984, newman1986,ns-dust,nm2002} and for $n\ge 5$~\cite{gj2004}.
They showed that the singularity formed is censored for $n(\ge 6)$, while it is naked for $n=4$.
For $n=5$, the singularity can be censored depending on the parameters in the initial data.

This paper is organized as follows. 
In Sec.~II, for the $n(\ge 5)$-dimensional spacetime $M \approx M^2 \times K^{n-2}$ with a perfect fluid and a cosmological constant, where $K^{n-2}$ is the $(n-2)$-dimensional Einstein space, we define a scalar on $M^2$, of which dimension is mass, and give a simple formulation of the basic equations in Einstein-Gauss-Bonnet gravity.
In Sec.~III, using this formalism, we investigate the final fate of the $n(\ge 5)$-dimensional spherically symmetric gravitational collapse of a dust cloud without a cosmological constant.
Section~V is devoted to discussion and conclusions.
In Appendix~A, we review the study of the general relativistic case for comparison and give some complements. 
Throughout this paper we use units such that $c=1$. 
As for notation we follow~\cite{Gravitation}. 
The Greek indices run $\mu=0,1, \cdots, n-1$.

\section{Model and basic equations}

We begin with the following $n$-dimensional ($n \geq 5$) action:
\begin{equation}
\label{action}
S=\int d^nx\sqrt{-g}\biggl[\frac{1}{2\kappa_n^2}(R-2\Lambda+\alpha{L}_{GB}) \biggr]+S_{\rm matter},
\end{equation}
where
$R$ and $\Lambda$ are the $n$-dimensional Ricci scalar and the cosmological constant, respectively. $\kappa_n\equiv\sqrt{8\pi G_n}$, where $G_n$ is the $n$-dimensional gravitational constant.
The Gauss-Bonnet term ${L}_{GB}$ is the combination of the Ricci scalar, Ricci tensor $R_{\mu\nu}$, and Riemann tensor $R^\mu_{~~\nu\rho\sigma}$ as
\begin{equation}
{L}_{GB}=R^2-4R_{\mu\nu}R^{\mu\nu}+R_{\mu\nu\rho\sigma}R^{\mu\nu\rho\sigma}.
\end{equation}
$\alpha$ is the coupling constant of the Gauss-Bonnet term. 
This type of action is derived in the low-energy limit of heterotic superstring theory~\cite{Gross}.
In that case, $\alpha$ is regarded as the inverse string tension and positive definite, and thus we assume $\alpha > 0$ in this paper.
We consider a perfect fluid as a matter field, whose action is represented by $S_{\rm matter}$ in Eq.~(\ref{action}).
We do not consider the case with $n \le 4$, in which the Gauss-Bonnet term does not contribute to the field equations.

The gravitational equation of the action (\ref{action}) is
\begin{equation}
{G}^\mu_{~~\nu} +\alpha {H}^\mu_{~~\nu} +\Lambda \delta^\mu_{~~\nu}= \kappa_n^2 {T}^\mu_{~~\nu}, \label{beq}
\end{equation}
where 
\begin{eqnarray}
{G}_{\mu\nu}&\equiv&R_{\mu\nu}-{1\over 2}g_{\mu\nu}R,\\
{H}_{\mu\nu}&\equiv&2\Bigl[RR_{\mu\nu}-2R_{\mu\alpha}R^\alpha_{~\nu}-2R^{\alpha\beta}R_{\mu\alpha\nu\beta}+R_{\mu}^{~\alpha\beta\gamma}R_{\nu\alpha\beta\gamma}\Bigr]
-{1\over 2}g_{\mu\nu}{L}_{GB}.
\end{eqnarray}
The energy-momentum tensor of a perfect fluid is
\begin{eqnarray}
{T}_{\mu\nu}=(p+\rho)u_{\mu}u_{\nu}+pg_{\mu\nu},
\end{eqnarray}
where $u^\mu$, $\rho$ and $p$ are the $n$-velocity of the fluid element, energy density, and pressure, respectively.
\begin{lm}
\label{em}
If $p=-\rho$, then $\rho$ is constant.
\end{lm}
{\it Proof.}
The energy-momentum conservation equation ${T}^{~\nu}_{\mu~;\nu}=0$ becomes
\begin{eqnarray}
\rho_{;\nu}u^{\nu} + (\rho + p) u^\mu_{~;\mu} &=&0, \label{b1}\\
(\rho +p) u_{\mu;\nu} u^{\nu} &=& -p_{,\nu}h^{\nu}_{~\mu},\label{b2}
\end{eqnarray}
where
$h_{\mu\nu} \equiv g_{\mu\nu}+u_\mu u_\nu$ is the projection tensor.
From Eqs.~(\ref{b1}) and (\ref{b2}), $\rho$ is constant if $p=-\rho$.
$\Box$
\vspace*{1cm}

By Lemma~\ref{em}, a perfect fluid obeying $p=-\rho$ is equivalent to a cosmological constant.
We assume $p \ne -\rho$ in this paper.

We consider the $n$-dimensional spacetime as a product manifold $M \approx M^2\times K^{n-2}$, where $K^{n-2}$ is the $(n-2)$-dimensional Einstein space, with the line element
\begin{equation}
\label{metric}
ds^2=-e^{2\Phi(t,r)}dt^2+e^{2\Psi(t,r)}dr^2+S(t,r)^2\gamma_{ij}dx^i dx^j,
\end{equation}  
where $\gamma_{ij}$ is the unit curvature metric on $K^{n-2}$.
Hereafter, a dot and a prime denote the differentiation with respect to $t$ and $r$, respectively. 
$k$ denotes the curvature of $K^{n-2}$ and takes $1$ (positive curvature), $0$ (zero curvature), and $-1$ (negative curvature).
We adopt the comoving coordinates such that the $n$-velocity of the fluid element is 
\begin{eqnarray}
u^\mu\frac{\partial}{\partial x^\mu}=e^{-\Phi}\frac{\partial}{\partial t}.
\end{eqnarray}  
The following Lemma is necessary to give our formalism of the basic equations.

\begin{lm}
\label{lm:tr}
If $p \ne -\rho$, then $S^2+2(n-3)(n-4)\alpha(k+e^{-2\Phi}\dot{S}^2-e^{-2\Psi}{S'}^2)$ cannot be zero.
\end{lm}
{\it Proof.}
If the relation
\begin{equation}
S^2+2(n-3)(n-4)\alpha(k+e^{-2\Phi}\dot{S}^2-e^{-2\Psi}{S'}^2)=0
\end{equation} 
is satisfied at a moment, then the ($t,t$) and ($r,r$) components of the field equation (\ref{beq}) give
\begin{eqnarray}
\frac{(n-1)(n-2)}{8\alpha(n-3)(n-4)}+\Lambda&=& -\kappa_n^2\rho \label{lmtr1}
\end{eqnarray} 
and 
\begin{eqnarray}  
\frac{(n-1)(n-2)}{8\alpha(n-3)(n-4)}+\Lambda&=& \kappa_n^2 p, \label{lmtr2}
\end{eqnarray}  
respectively.
Eqs.~(\ref{lmtr1}) and (\ref{lmtr2}) give a contradiction $p=-\rho$.
$\Box$
\vspace*{1cm}

Here we give a definition of a scalar on $M^2$ with the dimension of mass such that
\begin{eqnarray}
\label{qlm}
m &\equiv& \frac{(n-2)V_{n-2}^k}{2\kappa_n^2}[-{\tilde \Lambda}S^{n-1}+S^{n-3}(k-S_{,\mu}S^{,\mu})+{\tilde \alpha}S^{n-5}(k-S_{,\mu}S^{,\mu})^2],
\end{eqnarray}  
where ${\tilde \alpha}\equiv(n-3)(n-4)\alpha$, ${\tilde \Lambda}\equiv2\Lambda/[(n-1)(n-2)]$, and a comma denotes the partial differentiation.
A constant $V_{n-2}^k$ is the surface area of the $(n-2)$-dimensional unit Einstein space if it is compact.
For example, $V_{n-2}^1\equiv2\pi^{(n-1)/2}/\Gamma((n-1)/2)$ is the surface area of the $(n-2)$-dimensional unit sphere, where $\Gamma(x)$ is the gamma function.
In the four-dimensional spherically symmetric case without a cosmological constant, $m$ is reduced to the Misner-Sharp mass~\cite{ms1964}.

Then, the field equations are written in the following simple form:
\begin{eqnarray}
p'&=& -(\rho+p)\Phi', \label{basic1}\\
\dot{\rho}&=& -(\rho+p)\left[\dot{\Psi}+(n-2)\frac{\dot{S}}{S}\right], \label{basic2}\\
m' &=& V_{n-2}^k\rho S' S^{n-2},   \label{basic3}\\
\dot{m} &=& -V_{n-2}^kp \dot{S} S^{n-2}, \label{basic4}\\ 
0&=&-\dot{S'}+\Phi' \dot{S}+\dot{\Psi}S',\label{basic5}\\
m &=& \frac{(n-2)V_{n-2}^k}{2\kappa_n^2}[-{\tilde \Lambda}S^{n-1}+S^{n-3}(k+e^{-2\Phi}\dot{S}^2-e^{-2\Psi}{S'}^2)+{\tilde \alpha}S^{n-5}(k+e^{-2\Phi}\dot{S}^2-e^{-2\Psi}{S'}^2)^2].\label{basic6}
\end{eqnarray}  
The first two equations are the energy-momentum conservation equations. 
Eq.~(\ref{basic5}) is derived from the ($t,r$) component of Eq.~(\ref{beq}) with Lemma~\ref{lm:tr}.
Eq.~(\ref{basic6}) is obtained from Eq.~(\ref{qlm}).
Eqs.~(\ref{basic3}) and (\ref{basic4}) are obtained from the ($t,t$) and ($r,r$) components of Eq.~(\ref{beq}) by using Eqs.~(\ref{basic5}) and (\ref{basic6}).
Five of the above six equations are independent.

\section{Spherically symmetric dust cloud}
After this, we only consider the spherically symmetric collapse of a dust fluid without a cosmological constant, i.e., $p=0$ and $\Lambda=0$, for simplicity.
We assume the positive energy density, i.e., $\rho>0$.
Then Eq.~(\ref{basic1}) implies that $\Phi=\Phi(t)$, so that we can set $\Phi=0$ by redefinition of our time coordinate without a loss of generality.
Throughout this paper, we call the direction of increasing (decreasing) $t$ future (past).
Eq.~(\ref{basic4}) implies that $m=m(r)$, which is an arbitrary function.
$m$ is naturally interpreted as the mass inside the comoving radius $r$ because Eq.~(\ref{basic3}) implies 
\begin{eqnarray}
m = \int^rV_{n-2}^1\rho S^{n-2}\frac{\partial S}{\partial r}dr
\end{eqnarray}  
on the hypersurface with a constant $t$.
 From Eq.~(\ref{basic5}), we obtain
\begin{eqnarray}
e^{2\Psi}=\frac{{S'}^2}{1+f(r)},
\end{eqnarray}  
where $f(r)$ is an arbitrary function satisfying $f>-1$.
Eq.~(\ref{basic3}) gives
\begin{eqnarray}
\rho=\frac{m'}{V_{n-2}^1S'S^{n-2}}. \label{GBLTB-rho}
\end{eqnarray}  
From this equation, we find that there may exist both shell-crossing singularities, where $S'=0$, and shell-focusing singularities, where $S=0$.
Hereafter we assume $S'>0$, or equivalently $m'>0$, in order that shell-crossing singularities may be removed from our consideration.

From Eq.~(\ref{basic6}), we obtain a master equation of the system:
\begin{equation}
{\dot S}^2=f-\frac{S^2}{2{\tilde \alpha}}\left(1 \mp \sqrt{1+\frac{8{\tilde \alpha}\kappa_n^2m}{(n-2)V_{n-2}^1S^{n-1}}}\right). \label{dotS}
\end{equation}  
For later convenience, we call the solution of Eq.~(\ref{dotS}) the Gauss-Bonnet-Lema{\^ i}tre-Tolman-Bondi (GB-LTB) solution although the explicit form of the solution is not obtained in this paper. 
There are two families of solutions which correspond to the sign in front of the square root in Eq.~(\ref{dotS}).
We call the family with the minus (plus) sign the minus-branch (plus-branch) solution.
In the general relativistic limit ${\tilde \alpha} \to 0$, the minus-branch solution is reduced to the $n$-dimensional Lema{\^ i}tre-Tolman-Bondi solution.
The case with $n=4$ has been intensively studied by many authors~\cite{LTB,ns-dust,nm2002}, while the case with $n \ge 5$ has also been studied by several authors~\cite{gj2004}.
On the contrary, there is no general relativistic limit for the plus-branch solution.

We can consider that the GB-LTB solution is attached at a finite constant comoving radius $r=r_0>0$, where we represent this timelike hypersurface as $\Sigma$, to the outside vacuum region.
The outside vacuum region is represented by the solution independently discovered by Boulware and Deser~\cite{GB_BH} and Wheeler~\cite{Wheeler_1}, whose metric is 
\begin{eqnarray}
\label{f-eq}
ds^2=-F(S)dT^2+\frac{dS^2}{F(S)}+S^2d\Omega_{n-2}^2,
\end{eqnarray}  
where $d\Omega_{n-2}^2$ is the line element of the $(n-2)$-dimensional unit sphere.
$F(S)$ is defined by 
\begin{equation}
F(S) \equiv 1+\frac{S^2}{2\tilde{\alpha}}\Biggl(1\mp\sqrt{1+\frac{8{\tilde \alpha}\kappa_n^2M}{(n-2)V^1_{n-2} S^{n-1}}}\Biggr), \label{horizon}
\end{equation}  
where $M$ is a constant.
We call this solution the GB-Schwarzschild solution.
There are also two branches of the solution which correspond to the sign in front of the square root in Eq.~(\ref{horizon}).
The global structures of the GB-Schwarzschild solution are presented in~\cite{tm2005}.
The minus-branch GB-Schwarzschild solution is asymptotically flat, while the plus-branch solution is asymptotically anti-de~Sitter in spite of the absence of a cosmological constant.
The constant $M$ in the minus-branch GB-Schwarzschild solution is the higher-dimensional ADM mass~\cite{massGB}.

The position of $\Sigma$ and the relation between $T$ and $t$ on $\Sigma$ are represented by $S=S_\Sigma(t)$ and $T=T_\Sigma(t)$, respectively.
Now we derive the governing equations for $T_\Sigma$ and $S_\Sigma$~\cite{Poisson}.

As seen from inside of $\Sigma$, the metric on $\Sigma$ is obtained by
\begin{equation}
\label{1stin}
ds_{\Sigma}^2=-dt^2+S_{\Sigma}^2d\Omega^2_{n-2}.
\end{equation}
As seen from outside of $\Sigma$, on the other hand, it is 
\begin{equation}
\label{1stout}
ds_{\Sigma}^2=-\left(F(S_{\Sigma}){\dot T}_{\Sigma}^2-\frac{{\dot S}_{\Sigma}^2}{F(S_{\Sigma})}\right)dt^2+S_{\Sigma}^2d\Omega^2_{n-2}.
\end{equation}
Because the induced metric must be the same on both sides of the hypersurface $\Sigma$, we have
\begin{eqnarray}
1=F(S_{\Sigma}){\dot T}_{\Sigma}^2-\frac{{\dot S}_{\Sigma}^2}{F(S_{\Sigma})}.\label{cond1st1}
\end{eqnarray}
Here we define $\beta$ by
\begin{eqnarray}
\label{beta}
\beta \equiv \sqrt{F(S_{\Sigma})+{\dot S}_{\Sigma}^2}=F(S_{\Sigma}){\dot T}_{\Sigma}.
\end{eqnarray}

The unit normal $n^\mu$ to $\Sigma$ can be obtained from the relations $n_\mu u^\mu=0$ and $n_\mu n^\mu=1$.
As seen from inside $\Sigma$, we obtain 
\begin{eqnarray}
u^\mu\frac{\partial}{\partial x^\mu}=\frac{\partial}{\partial t}
\end{eqnarray}
and 
\begin{eqnarray}
n_\mu dx^\mu =\frac{S'}{\sqrt{1+f(r_0)}}dr,
\end{eqnarray}
where we have chosen $n^r>0$ so that $n^\mu$ is directed toward $\Sigma$.
As seen from outside $\Sigma$, we obtain 
\begin{eqnarray}
u^\mu\frac{\partial}{\partial x^\mu}={\dot T}_{\Sigma}\frac{\partial}{\partial t}+{\dot S}_{\Sigma}\frac{\partial}{\partial r}
\end{eqnarray}
and 
\begin{eqnarray}
n_\mu dx^\mu =-{\dot S}_{\Sigma} dt+{\dot T}_{\Sigma}dr
\end{eqnarray}
with a consistent choice for the sign.

As seen from inside of $\Sigma$, the non-zero components of the extrinsic curvature $K^a_{~b}$ of $\Sigma$ are calculated as
\begin{eqnarray}
K^t_{~t}=0 \label{2ndin1}
\end{eqnarray}
and 
\begin{eqnarray}
K^i_{~i}=\frac{\sqrt{1+f(r_0)}}{S_{\Sigma}}. \label{2ndin2}
\end{eqnarray}
As seen from outside of $\Sigma$, we obtain
\begin{equation}
K^t_{~t}=\frac{{\dot \beta}}{{\dot S}_{\Sigma}}\left(F(S_{\Sigma}){\dot T}_{\Sigma}^2-\frac{{\dot S}_{\Sigma}^2}{F(S_{\Sigma})}\right)^{-1} \label{2ndout1}
\end{equation}
and
\begin{eqnarray}
K^i_{~i}=\frac{\beta}{S_{\Sigma}}. \label{2ndout2}
\end{eqnarray}

In order to have a smooth transition at $\Sigma$, we demand that $K^a_{~b}$ be the same on both sides of $\Sigma$.
From Eqs.~(\ref{2ndin2}) and (\ref{2ndout2}), we obtain 
\begin{eqnarray}
\beta=\sqrt{1+f(r_0)},
\end{eqnarray}
which is consistent with Eqs.~(\ref{2ndin1}) and (\ref{2ndout1}).

From Eq.~(\ref{beta}), we finally obtain the equations of motion for the hypersurface $\Sigma$ as
\begin{eqnarray}
\frac{dT_{\Sigma}}{dt}&=&\frac{\sqrt{1+f(r_0)}}{F(S_{\Sigma})}, \label{s3} \\
\left(\frac{dS_{\Sigma}}{dt}\right)^2&=&1-F(S_{\Sigma})+f(r_0), \nonumber \\
&=&f(r_0)-\frac{S_{\Sigma}^2}{2\tilde{\alpha}}\Biggl(1\mp\sqrt{1+\frac{8{\tilde \alpha}\kappa_n^2M}{(n-2)V^1_{n-2} S_{\Sigma}^{n-1}}}\Biggr). \label{s2}
\end{eqnarray}

The following Lemma shows that $m$ is a well-defined quasi-local mass:
\begin{lm}
\label{lm:mass}
The GB-LTB solution can be attached at $r=r_0$ to the outside GB-Schwarzschild solution only in the same branch and then $M=m(r_0)$.
\end{lm}
{\it Proof.}
From Eq.~(\ref{dotS}), we also obtain ${\dot S}_{\Sigma}^2$ as
\begin{equation}
\label{s1}
\left(\frac{dS_{\Sigma}}{dt}\right)^2=f(r_0)-\frac{S_{\Sigma}^2}{2{\tilde \alpha}}\left(1\mp\sqrt{1+\frac{8{\tilde \alpha}\kappa_n^2m(r_0)}{(n-2)V_{n-2}^1S_{\Sigma}^{n-1}}}\right).
\end{equation}
From Eqs.~(\ref{s2}) and (\ref{s1}), it is found that only the solution in the same branch can be attached and then 
\begin{eqnarray}
M=m(r_0)
\end{eqnarray}
is satisfied.  
$\Box$

\section{Final fate of gravitational collapse}
In this section, we consider the final fate of gravitational collapse.
First we consider the possibility of bounce.  
We define $v(t,r)$ as
\begin{eqnarray}  
v \equiv \frac{S}{r}, \label{def-v}
\end{eqnarray}
from which we find that ${\dot v}<(>)0$ is equivalent to ${\dot S}<(>)0$.
We set
\begin{eqnarray}  
v(t_{\rm i},r)=1 \label{initialtime}
\end{eqnarray}
by using the freedom of the radial coordinate such as ${\bar r} \equiv {\bar r}(r)$, where $t_{\rm i}$ is the initial time.
Then, we have $S=r$ at the initial moment and $v=0$ at the singularity.
We assume that each shell collapses initially, i.e., 
\begin{eqnarray}  
{\dot v}(t_{\rm i},r)<0.
\end{eqnarray}
We represent the time when the singularity occurs for each $r$ as $t_{\rm s}(r)$, i.e., 
\begin{eqnarray}  
v(t_{\rm s}(r),r)=0.
\end{eqnarray}
If ${\dot v}=0$ is satisfied for some $r$ after the initial moment, the shell ceases to collapse and then bounces (${\dot v}>0$).

We define ${\cal M}(r)$ and $b(r)$ such as
\begin{eqnarray}  
\frac{2\kappa_n^2}{(n-2)V^1_{n-2}}m(r)&=&r^{n-1}{\cal M}(r), \label{def-M}\\
f(r)&=&r^2b(r),\label{def-b}
\end{eqnarray}
where ${\cal M}_0 \equiv {\cal M}(0)<\infty$ and $b_0 \equiv b(0)<\infty$ are satisfied for the regularity at the center.
By the following Lemma, shell-focusing singularities are never formed in the plus-branch solution for $n \ge 6$.
\begin{lm}
\label{lm:bounce}
Bounce inevitably occurs in the plus-branch solution for $n \ge 6$.
\end{lm}
{\it Proof.}
Eq.~(\ref{dotS}) gives
\begin{eqnarray}
{\dot v}^2&=&b-\frac{v^2}{2{\tilde \alpha}}\left(1 \mp \sqrt{1+\frac{4{\tilde \alpha}{\cal M}}{v^{n-1}}}\right), \label{master-zero}\\
&\equiv&G(v),
\end{eqnarray}  
where $G(v) \ge 0$ must be satisfied.
$v=0$ cannot be achieved in the plus-branch solution for $n \ge 6$ because 
\begin{eqnarray}
\lim_{v \to 0}G(v)=-\infty
\end{eqnarray}  
is satisfied in that case.
$\Box$
\vspace*{1cm}

We obtain
\begin{equation}
\frac{dG(v)}{dv} = \frac{\pm v^{n-1}\left(1 \mp \sqrt{1+4{\tilde \alpha}{\cal M}/v^{n-1}}\right)\mp(n-5){\tilde \alpha}{\cal M}}{{\tilde \alpha}v^{n-2}\sqrt{1+4{\tilde \alpha}{\cal M}/v^{n-1}}},
\end{equation}  
which reads that $G(v)$ is a decreasing function for the minus-branch solution and the plus-branch solution with $n=5$.
Since 
\begin{eqnarray}
G(1)=b-\frac{1}{2{\tilde \alpha}}\left(1 \mp \sqrt{1+4{\tilde \alpha}{\cal M}}\right)
\end{eqnarray}  
is obtained, we assume
\begin{eqnarray}
G(1)=b(r)-\frac{1}{2{\tilde \alpha}}\left(1 \mp \sqrt{1+4{\tilde \alpha}{\cal M}(r)}\right) \ge 0 \label{nobounce}
\end{eqnarray}  
holds for $0 \le r \le r_0$.

Now we can show the following theorem:
\begin{The}
\label{the:cch0}
Let us consider the $n(\ge 5)$-dimensional gravitational collapse of a spherical dust cloud with positive mass in Einstein-Gauss-Bonnet gravity without a cosmological constant.
Then, the singularity formed in the outside vacuum region is (i) timelike in the five-dimensional plus-branch and in the five-dimensional minus-branch with $0<m(r_0)<3{\tilde\alpha}V_3^1/(2\kappa^2_5)$
(ii) outgoing-null in the five-dimensional minus-branch with $m(r_0)=3{\tilde\alpha}V_3^1/(2\kappa^2_5)$
(iii) spacelike in the $n(\ge 6)$-dimensional minus-branch and in the five-dimensional minus-branch with $m(r_0)>3{\tilde\alpha}V_3^1/(2\kappa^2_5)$.
\end{The}
{\it Proof.}
Trivial from Lemma~\ref{lm:mass}, Lemma~\ref{lm:bounce} and the result in~\cite{tm2005}.
$\Box$
\vspace*{1cm}

From Theorem~\ref{the:cch0}, the following corollary holds:
\begin{Coro}
\label{cr:cch0}
Let us consider the five-dimensional gravitational collapse of a spherical dust cloud with positive mass in Einstein-Gauss-Bonnet gravity without a cosmological constant.
Then, the weak cosmic censorship hypothesis is violated in the plus-branch and in the minus-branch with $0<m(r_0)<3{\tilde\alpha}V_3^1/(2\kappa^2_5)$.
\end{Coro}
\vspace*{1cm}

Next let us consider the trapped surfaces in the collapsing solution.
In order to define the trapped surface, we consider the time evolution of the areal radius along a future-directed radial null geodesic, which we denote $dS/dt|_{+}$ and $dS/dt|_{-}$ for outgoing and ingoing null geodesics, respectively.
Future-directed radial null geodesics obey
\begin{equation}
\frac{dr}{dt}\biggl|_{\pm}=\pm \frac{\sqrt{1+f}}{S'}, \label{null}
\end{equation}  
where the signs $+$ and $-$ correspond to the outgoing and ingoing null geodesics, respectively.
Using Eq.~(\ref{null}) with ${\dot S}<0$ and $S'>0$, we obtain
\begin{eqnarray}
\frac{dS}{dt}\biggl|_{+}&\equiv&{\dot S}+S'\frac{dr}{dt}\biggl|_{+} ,\nonumber \\
  &=&-\left[f-\frac{S^2}{2{\tilde \alpha}}\left(1 \mp \sqrt{1+\frac{8{\tilde \alpha}\kappa_n^2m}{(n-2)V_{n-2}^1S^{n-1}}}\right)\right]^{1/2}+\sqrt{1+f}, \label{expansion} \\
\frac{dS}{dt}\biggl|_{-}&\equiv&{\dot S}+S'\frac{dr}{dt}\biggl|_{-} ,\nonumber \\
  &=&-\left[f-\frac{S^2}{2{\tilde \alpha}}\left(1 \mp \sqrt{1+\frac{8{\tilde \alpha}\kappa_n^2m}{(n-2)V_{n-2}^1S^{n-1}}}\right)\right]^{1/2}-\sqrt{1+f}, \label{expansion-}
\end{eqnarray}
where the signs $-$ and $+$ in front of the square root in Eqs.~(\ref{expansion}) and (\ref{expansion-}) correspond to the minus-branch and the plus-branch solutions, respectively.
\begin{dn}
A trapped surface is a metric sphere with $dS/dt|_{+}< 0$ and $dS/dt|_{-}< 0$.
\label{dn:trap}
\end{dn}
\begin{dn}
A trapped region is a region where trapped surfaces exist.
\label{dn:tr}
\end{dn}
\begin{dn}
An apparent horizon is a boundary of a trapped region.
\label{dn:ah}
\end{dn}
\begin{lm}
$dS/dt|_{-}<0$ holds in the collapsing GB-LTB solution.
\label{lm:trap-}
\end{lm}
{\it Proof.}
Trivial from Eq.~(\ref{expansion-}).
$\Box$
\begin{lm}
There is no trapped surface in the collapsing plus-branch GB-LTB solution.
\label{lm:trap+}
\end{lm}
{\it Proof.}
From Eq.~(\ref{expansion}), $dS/dt|_{+} < 0$ is equivalent to 
\begin{equation}
1+\frac{S^2}{2\tilde{\alpha}} < -\frac{S^2}{2\tilde{\alpha}}\sqrt{1+\frac{8{\tilde \alpha}\kappa_n^2m}{(n-2)V^1_{n-2} S^{n-1}}}
\end{equation}
in the plus-branch solution, which cannot be satisfied because the left-hand-side is positive definite while the right-hand-side is negative definite.
$\Box$
\vspace*{1cm}

By Lemma~\ref{lm:bounce}, singularities are never formed in the plus-branch solution for $n \ge 6$.
In the plus-branch solution when $n=5$, the singularity formed should be naked by Lemma~\ref{lm:trap+}.  
Hereafter we concentrate on the minus-branch solution.

\begin{lm}
In the collapsing minus-branch GB-LTB solution, a trapped region is obtained by
\begin{eqnarray}
m > \frac{(n-2)V_{n-2}^1}{2\kappa^2_n}(S^{n-3}+{\tilde\alpha}S^{n-5}), \label{trap}
\end{eqnarray}
while an apparent horizon is obtained by
\begin{eqnarray}
m =\frac{(n-2)V_{n-2}^1}{2\kappa^2_n}(S^{n-3}+{\tilde\alpha}S^{n-5}). \label{ah}
\end{eqnarray}
\end{lm}
{\it Proof.}
From Eq.~(\ref{expansion}), $dS/dt|_{+} \le 0$ is equivalent to 
\begin{equation}
F(t,r) \equiv 1+\frac{S^2}{2\tilde{\alpha}}\Biggl(1-\sqrt{1+\frac{8{\tilde \alpha}\kappa_n^2m}{(n-2)V^1_{n-2} S^{n-1}}}\Biggr)\le 0 \label{f-func}
\end{equation}  
in the minus-branch solution.
This is equivalent to 
\begin{eqnarray}
m \ge \frac{(n-2)V_{n-2}^1}{2\kappa^2_n}[S^{n-3}+{\tilde\alpha}S^{n-5}]. 
\end{eqnarray}  
$\Box$
\vspace*{1cm}

We represent the time when each shell reaches the singularity as $t=t_{s}(r)$.
The regular region of spacetime is $t < t_s(r)$. 
The following Lemma will be used later.
\begin{lm}
\label{lm:sing}
$t=t_{s}(r)$ is a non-decreasing function.
\end{lm}
{\it Proof.}
If $t=t_s(r)$ has a decreasing portion, then we can find constants $t_1$, $t_2$, $r_1$, and $r_2$ such that $t_1<t_2$ and $0<r_1<r_2$ with 
\begin{eqnarray}
S(t_1,r_2)=S(t_2,r_1)=0. \label{condition1}
\end{eqnarray}  
From the assumptions $S'>0$ and ${\dot S}<0$, we obtain $S(t_1,r_2)>S(t_1,r_1)$ and $S(t_2,r_1)<S(t_1,r_1)$, respectively.
These two inequalities give
\begin{eqnarray}
S(t_1,r_2)>S(t_2,r_1),
\end{eqnarray}
which contradicts Eq.~(\ref{condition1}).
$\Box$
\vspace*{1cm}

We consider whether the singularity is naked or censored.
Future-directed ingoing geodesics, which satisfy $dt/dr<0$, cannot emanate from the singularity by Lemma~\ref{lm:sing}, while future-directed outgoing geodesics, which satisfy $dt/dr>0$, may do.
Lemma~\ref{lm:trap-} and the contraposition of the following Lemma imply that it is sufficient to consider only the future-directed outgoing radial null geodesics to find whether the singularity is naked or censored.
The proof is similar to the four-dimensional case in~\cite{nmg2002}.
\begin{lm}
\label{lm:geodesics}
If a future-directed outgoing causal (excluding radial null) geodesic emanates from the singularity, then a future-directed outgoing radial null geodesic emanates from the singularity.
\end{lm}
{\it Proof.}
In the spacetime (\ref{metric}), the tangent to a causal geodesic satisfies
\begin{eqnarray}
-e^{2\Phi}\left(\frac{dt}{d\lambda}\right)^2+e^{2\Psi}\left(\frac{dr}{d\lambda}\right)^2+\frac{L^2}{S^2}=\epsilon, \label{hamilton}
\end{eqnarray}  
where $\lambda$ is an affine parameter, $L$ is the conserved angular momentum, and $\epsilon=0,-1$ for null and timelike geodesics, respectively.
Then, at any point on such a geodesic,
\begin{eqnarray}
e^{2\Phi}\left(\frac{dt}{d\lambda}\right)^2 \ge e^{2\Psi}\left(\frac{dr}{d\lambda}\right)^2
\end{eqnarray}
with equality holding only for radial null geodesics.
On the $(r,t)$-plane, this gives
\begin{eqnarray}
\frac{dt}{dr}\ge e^{\Psi-\Phi},
\end{eqnarray}
where we take the positive root for the future-directed geodesics.
This gives
\begin{eqnarray}
\frac{dt_{\rm CG}}{dr}>\frac{dt_{\rm RNG}}{dr}, \label{geodesics1}
\end{eqnarray}
where the subscripts represent causal (excluding radial null) geodesics and outgoing radial null geodesics, respectively.
Now suppose that $t=t_{\rm CG}(r)$ extends back to a singularity located at $(r,t)=(r_s,t_{s}(r_s))$.
Let $p$ be any point on $t=t_{\rm CG}(r)$ to the future of the singularity.
Applying inequality (\ref{geodesics1}) at $p$, we see that the $t=t_{\rm RNG}(r)$ through $p$ crosses $t=t_{\rm CG}(r)$ from above and hence points $t_{\rm RNG}(r)$ on this radial null geodesic prior to $p$ must lie to the future of points on $t=t_{\rm CG}(r)$ prior to $p$, in the sense of $t_{\rm RNG}(r)>t_{\rm CG}(r)$ for $r \in (r_s,r_{\ast})$, where $r_{\ast}$ corresponds to $p$.
Thus, the radial null geodesics, which necessarily lie at $t<t_{s}(r)$, must extend back to the singularity at $r \in [r_s,r_{\ast})$, and so must emerge from the singularity.
$\Box$
\begin{The}
In the collapsing minus-branch GB-LTB solution, massive singularities for $n \ge 6$ and singularities with $m>3{\tilde\alpha}V_{3}^1/(2\kappa^2_5)$ for $n=5$ are censored.
\label{lm:cch1}
\end{The}
{\it Proof.}
Eq.~(\ref{trap}) gives that the central singularities with $m>0$ for $n \ge 6$ and $m>3{\tilde\alpha}V_{3}^1/(2\kappa^2_5)$ for $n=5$ are in the trapped region.
By Lemma~\ref{lm:trap-} and the contraposition of Lemma~\ref{lm:geodesics}, it is sufficient to show that the future-directed outgoing radial null geodesics cannot emerge from the singularity.
Past-directed ingoing radial null geodesics emanating from a spacetime event in the trapped region cannot reach the singularity at $S=0$ because $dS/dt|_{+}<0$ holds there.
Therefore, there is no future-directed radial null geodesic emanating from the singularity in the trapped region.
$\Box$
\vspace*{1cm}

By Theorem~\ref{lm:cch1}, if naked singularities are formed in the minus-branch GB-LTB solution, they are massless for $n \ge 6$, while they have mass with $0 \le m \le 3{\tilde\alpha}V_{3}^1/(2\kappa^2_5)$ for $n=5$.
In the next two subsections, we investigate the final fate of the gravitational collapse further.
First, we consider the homogeneous collapse with $f=0$ as the simplest case; subsequently we move to the inhomogeneous case.

\subsection{Homogeneous collapse}
Here we consider homogeneous collapse with $f=0$, which is represented by the flat Friedmann-Robertson-Walker (FRW) solution.
Then, we have $S=ra(t)$ and 
\begin{eqnarray}
ds^2=-dt^2+a(t)^2(dr^2+r^2d\Omega_{n-2}^2), 
\end{eqnarray}  
where $a(t)$ is the scale factor.
From Eq.~(\ref{master-zero}), we find that only the minus-branch is possible and ${\cal M}={\cal M}_0$, where ${\cal M}_0$ is a positive constant.
Eq.~(\ref{master-zero}) is then reduced to
\begin{eqnarray}
{\cal M}_0=a^{n-3}\dot{a}^2+{\tilde \alpha}a^{n-5}\dot{a}^4. \label{F-eq}
\end{eqnarray}  
\begin{The}
\label{the:homo}
In the collapsing flat FRW solution with a dust fluid in Einstein-Gauss-Bonnet gravity, the big-crunch singularity is spacelike for $n \ge 6$, while it is ingoing-null for $n=5$.
\end{The}
{\it Proof.}
From Eq.~(\ref{F-eq}), we find that the scale factor $a$ behaves as
\begin{eqnarray}
a \simeq (-t)^{4/(n-1)} \label{a-flatFRW}
\end{eqnarray}
near $t=0$, where we set the origin of $t$ corresponding to $a=0$.
The big-crunch singularity is formed at $t=0$, where the Kretschmann invariant $K \equiv R_{\mu\nu\rho\sigma}R^{\mu\nu\rho\sigma}$ diverges as 
\begin{eqnarray}
K={\cal O}(1/(-t)^4).
\end{eqnarray}  

The structure of the singularity is determined by Eq.~(\ref{a-flatFRW}).
We take the line element of the flat FRW solution with the scale factor $a=(-t/t_0
)^p$ to the conformally flat form as
\begin{eqnarray}
ds^2=a(t(\eta))^2(-d\eta^2+dr^2+r^2d\Omega_{n-2}^2),
\end{eqnarray}  
where $d\eta \equiv dt/a(t)$ and $t_0$ is a constant.
The range of $\eta$ is $-\infty<\eta<\infty$ and $-\infty<\eta<\eta_0$ for $p=1$ and $0<p<1$, respectively, where $\eta_0$ is a constant. 
Thus, the singularity is spacelike for $0<p<1$, while it is ingoing-null for $p=1$~\cite{senovilla}.
$\Box$
\vspace*{1cm}

\begin{figure}[tbp]
\includegraphics[width=.50\linewidth]{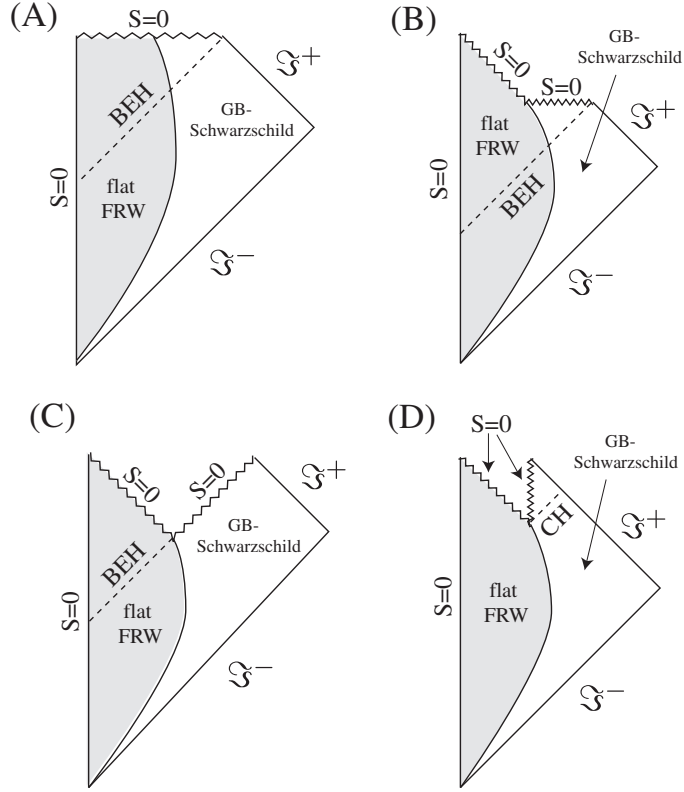}
\caption{
Global structures of the $n(\ge 5)$-dimensional homogeneous collapse of a spherically symmetric dust cloud with $f=0$.
Zigzag lines represent the central singularities. 
$\Im^{+(-)}$ corresponds to the future (past) null infinity. 
BEH and CH stand for the black-hole event horizon and the Cauchy horizon.
The global structure is represented by (A) for $n\ge 6$. 
For $n=5$, the global structures are (B), (C), and (D) which correspond to $m(r_0)>3{\tilde\alpha}V_3^1/(2\kappa^2_5)$, $m(r_0)=3{\tilde\alpha}V_3^1/(2\kappa^2_5)$, and $m(r_0)<3{\tilde\alpha}V_3^1/(2\kappa^2_5)$, respectively.
(A), (B), and (C) represent black-hole formation, while (D) represents a globally naked singularity formation.
In general relativity, the global structure is represented by (A) for $n \ge 4$.
}
\label{Fig4}
\end{figure}

By Theorems~\ref{the:cch0} and \ref{the:homo}, global structures of the homogenous collapse of a dust cloud with $f=0$ are shown in Fig.~\ref{Fig4}.
In general relativity, this is the $n$-dimensional Oppenheimer-Snyder solution, which represents black-hole formation for $n \ge 4$.
In Einstein-Gauss-Bonnet gravity, the solution also represents black-hole formation for $n \ge 6$ and for $n=5$ with $m(r_0)\ge 3{\tilde\alpha}V_3^1/(2\kappa^2_5)$.
For $n=5$ with $m(r_0)< 3{\tilde\alpha}V_3^1/(2\kappa^2_5)$, on the other hand, the solution represents globally naked singularity formation.

\subsection{Inhomogeneous collapse}
Next we consider inhomogeneous collapse.
First we consider the distribution of the apparent horizon around $r=0$.
From Eq.~(\ref{trap}), the trapped region is given by
\begin{eqnarray}
{\cal M}>r^{-2}v^{n-3}+{\tilde \alpha}r^{-4}v^{n-5}. \label{trap2}
\end{eqnarray} 
From Eq.~(\ref{ah}), the apparent horizon is given by
\begin{eqnarray}
{\cal M}=r^{-2}v^{n-3}+{\tilde \alpha}r^{-4}v^{n-5}. \label{ah2}
\end{eqnarray} 
By Theorem~\ref{lm:cch1}, only the singularity at $r=0$ may be naked for $n \ge 6$.
On the other hand, the singularity at $r=r_s$ with $0 \le r_s \le r_{{\rm ah}}$ may be naked for $n=5$, where $r_{{\rm ah}}$ is defined by
\begin{eqnarray}
r_{\rm ah}^{4} {\cal M}(r_{\rm ah})={\tilde \alpha}. \label{rah}
\end{eqnarray}
From Eqs.~(\ref{trap2})-(\ref{rah}), we obtain
\begin{eqnarray}
\label{cr:cch0}
r_s^{4}{\cal M}(r_s) \le {\tilde \alpha}
\end{eqnarray}
for $n=5$.

Next we derive the relation between the time when the singularity occurs and the time when the apparent horizon appears for each $r$.
Here we write the master equation (\ref{master-zero}) again:
\begin{equation}  
{\dot v}=-\left[b-\frac{v^2}{2{\tilde \alpha}}\left(1-\sqrt{1+4{\tilde\alpha}{\cal M}v^{1-n}}\right)\right]^{1/2}.\label{vdot}
\end{equation}
Integrating this equation with respect to $v$, we obtain
\begin{equation}  
t(v,r)=\int^1_v\frac{dv}{\left[b-v^2\left(1-\sqrt{1+4{\tilde\alpha}{\cal M}v^{1-n}}\right)/(2{\tilde \alpha})\right]^{1/2}}+{\tilde t}(r), \label{t(vr)}
\end{equation}
where $r$ is to be treated as a constant in the above equation and ${\tilde t}$ is an arbitrary function of $r$.
We find ${\tilde t}(r)=t(1,r)$, so that ${\tilde t}(r)$ means the time for each $r$ when the areal radius $S$ coincides with the comoving radius $r$. 
We can set ${\tilde t}(r)=t_{\rm i}$, where $t_{\rm i}$ is a constant, by using the freedom of the radial coordinate such as ${\bar r} \equiv {\bar r}(r)$.
This means that $S=r$ is satisfied at a moment $t=t_{\rm i}$, which we have defined as the initial time in Eq.~(\ref{initialtime}).
Moreover, we can set $t_{\rm i}=0$ by using the freedom of the time translation.
Thus, we set ${\tilde t}(r) \equiv 0$ hereafter.

The time when each shell reaches the singularity is given from Eq.~(\ref{t(vr)}) by
\begin{equation}  
t_s(r) = \int^1_0\frac{dv}{\left[b-v^2\left(1-\sqrt{1+4{\tilde\alpha}{\cal M}v^{1-n}}\right)/(2{\tilde \alpha})\right]^{1/2}}. \label{ts-0}
\end{equation}
From Eq.~(\ref{t(vr)}), we also obtain the time $t=t_{\rm ah}(r)$ when the apparent horizon appears for each $r$ by
\begin{eqnarray}  
t_{\rm ah}(r)&\equiv&\int^1_{v_{\rm ah}(r)}\frac{dv}{\left[b-v^2\left(1-\sqrt{1+4{\tilde\alpha}{\cal M}v^{1-n}}\right)/(2{\tilde \alpha})\right]^{1/2}}, \nonumber \\
&=&\int^1_0\frac{dv}{\left[b-v^2\left(1-\sqrt{1+4{\tilde\alpha}{\cal M}v^{1-n}}\right)/(2{\tilde \alpha})\right]^{1/2}}+\int^0_{v_{\rm ah}(r)}\frac{dv}{\left[b-v^2\left(1-\sqrt{1+4{\tilde\alpha}{\cal M}v^{1-n}}\right)/(2{\tilde \alpha})\right]^{1/2}}, \nonumber \\
&=&t_{s}(r)-\int^{v_{\rm ah}(r)}_0\frac{dv}{\left[b-v^2\left(1-\sqrt{1+4{\tilde\alpha}{\cal M}v^{1-n}}\right)/(2{\tilde \alpha})\right]^{1/2}}, \label{tah1}
\end{eqnarray}
where $v=v_{\rm ah}(r)$ is given by solving Eq.~(\ref{ah2}) algebraically.

In the next two subsections, we consider whether the singularity appearing is naked or censored.
We treat the singularities at $r=0$ in the case of $n \ge 5$ and at $r=r_s$ in the case of $n=5$, separately.

\subsubsection{Singularities at $r=0$ for $n \ge 5$}
We consider not only regular but also smooth initial data at the symmetric center such as
\begin{eqnarray}  
{\cal M}(r)&=&{\cal M}_0+{\cal M}_2r^2+{\cal M}_4r^4+\cdots,\label{seriesm}\\
b(r)&=&b_0+b_2r^2+b_4r^4+\cdots ,\label{seriesb}
\end{eqnarray}
where ${\cal M}_2, {\cal M}_4, \cdots$ and $b_2, b_4, \cdots$ are constants.
Expanding Eq.~(\ref{t(vr)}) around $r=0$, we obtain
\begin{equation}  
t(v,r)=t(v,0)+\frac{r^2}{2}\chi_2(v)+\cdots, \label{texp}
\end{equation}
where $\chi_2(v)$ is defined by
\begin{eqnarray}  
\chi_2(v) \equiv -\int^1_v\left[b_0-\frac{v^2}{2{\tilde \alpha}}\left(1-\sqrt{1+4{\tilde\alpha}{\cal M}_0v^{1-n}}\right)\right]^{-3/2}\left(b_2+\frac{{\cal M}_2v^{3-n}}{\sqrt{1+4{\tilde\alpha}{\cal M}_0v^{1-n}}}\right)dv,
\end{eqnarray}
where the integrand is finite for $0 \le v \le 1$.
The time when the central shell reaches the singularity is given from Eq.~(\ref{ts-0}) by
\begin{equation}  
t_{s0}=\int^1_0\frac{dv}{\left[b_0-v^2\left(1-\sqrt{1+4{\tilde\alpha}{\cal M}_0v^{1-n}}\right)/(2{\tilde \alpha})\right]^{1/2}}. \label{ts0}
\end{equation}
From Eqs.~(\ref{texp}) and (\ref{ts0}), the time when other shells close to $r=0$ reach the singularity is given by
\begin{eqnarray}  
t_s(r) =t_{s0}+\frac{r^2}{2}\chi_2(0)+\cdots. \label{ts}
\end{eqnarray}
By Lemma~\ref{lm:sing}, $\chi_2(0)$ is non-negative and we assume $\chi_2(0) > 0$ in this paper.
From Eqs.~(\ref{t(vr)}) and (\ref{texp}) and the relation
\begin{eqnarray}  
0=\left(\frac{\partial t}{\partial v}\right)_r\left(\frac{\partial v}{\partial r}\right)_t+\left(\frac{\partial t}{\partial r}\right)_v, \label{dtdr}
\end{eqnarray}
we obtain 
\begin{equation}  
v'=\left[b_0-\frac{v^2}{2{\tilde \alpha}}\left(1-\sqrt{1+4{\tilde\alpha}{\cal M}_0v^{1-n}}\right)\right]^{1/2}(r\chi_2(v)+\cdots). \label{v'}
\end{equation}

Now we can show the following theorem.
\begin{The}
Let us consider the gravitational collapse of a spherical dust cloud with positive mass and smooth initial profiles satisfying $\chi_2(0) > 0$ in Einstein-Gauss-Bonnet gravity without a cosmological constant.
Then, the strong cosmic censorship hypothesis holds in the minus-branch for $n \ge 10$ and $n=9$ with $0<\chi_2(0)<({\cal M}_0/{\tilde\alpha})^{1/4}$. 
\end{The}
{\it Proof.}
By Theorems~\ref{the:cch0} and \ref{lm:cch1}, only the singularity at $r=0$ has the possibility of being naked for $n \ge 6$.  
A decreasing apparent horizon in the $(r,t)$-plane is a sufficient condition for the singularity formed to be censored because it shows the entrapment of the neighborhood of the center before the singularity.
Expanding the integrand in Eq.~(\ref{tah1}) in a power series in $v$ and keeping only the leading order term, we obtain
\begin{eqnarray}  
t_{\rm ah}(r)&=&t_{s0}+\frac{r^2}{2}\chi_2(0) \nonumber \\
&&+\cdots -\frac{4}{n-1}\left(\frac{{\cal M}_0}{{\tilde\alpha}}\right)^{1/(n-5)}r^{(n-1)/(n-5)}
\end{eqnarray}
for $n \ge 6$.
For $n \ge 10$, the last term dominates the second term and consequently the apparent horizon is decreasing near $r=0$ in the $(r,t)$-plane.
For $n=9$ with $0<\chi_2(0)<({\cal M}_0/{\tilde\alpha})^{1/4}$, the apparent horizon is also decreasing near $r=0$ in the $(r,t)$-plane.
$\Box$
\vspace*{1cm}

Next we consider the nakedness of the singularity for $5 \le n \le 9$.
By the contraposition of Lemma~\ref{lm:geodesics}, we consider only future-directed outgoing radial null geodesics.
We will show the existence of future-directed outgoing null geodesics emerging from the singularity at $r=r_s \ge 0$.
For this purpose, we adopt the fixed-point method~\cite{christodoulou1984,newman1986,khi2000}. 
Here we define
\begin{eqnarray}  
x \equiv \frac{S}{(r-r_s)^{q}}=\frac{rv}{(r-r_s)^{q}}, \label{x}
\end{eqnarray}
where $q(>0)$ is a constant and we consider only the region of $r \ge r_s$.
The equation for the future-directed outgoing radial null geodesics (\ref{null}) is written by $x$ as
\begin{eqnarray}
\frac{dx}{dr}+\frac{qx}{r-r_s}=\Xi(x,r),
\label{theta}
\end{eqnarray}
where $\Xi$ is defined by 
\begin{eqnarray}
\Xi(x,r) \equiv \frac{v+rv'}{(r-r_s)^q}\biggl(1+\frac{r{\dot v}}{\sqrt{1+r^2b}}\biggl). 
\end{eqnarray}
Problems appear in the null geodesic equation when we consider $r=r_s$.

We consider the neighborhood of $r=r_s$.
Now let us consider the case in which $\Xi$ is expanded around $r=r_s$ as
\begin{eqnarray}
\Xi(x,r) \simeq \frac{g_0(q)x+\sum^N_{i=1}g_i(q)x^{\alpha_i}}{r-r_s}+h(q,x)(r-r_s)^a+\textrm{o}((r-r_s)^{a}), \label{xi}
\end{eqnarray}
where $a(>-1)$ and $\alpha_i(\ne 1)$ are constants, $g_0$, $g_i$, and $h$ are some functions, and $N( \ge 1)$ is an integer.
Without a loss of generality, we assume $\alpha_i<\alpha_{i+1}$ for any $i$. 
We will use the following Lemma to prove the nakedness of the singularity.
\begin{lm}
\label{lm:fixedpoint}
If $\Xi(x,r)$ is expanded as Eq.~(\ref{xi}) with $N=1$ and $g_0(q) \ne q$, then there exists an asymptotic solution satisfying $x(0)=[g_1/(q-g_0)]^{1/(1-\alpha_1)}$ near $r=r_s$, and moreover it is the unique solution of Eq.~(\ref{theta}) which is continuous at $r=r_s$. 
\end{lm}
{\it Proof.}
By assumptions, the null geodesic equation (\ref{theta}) can be written near $r=r_s$ as
\begin{eqnarray}
\frac{dx}{dr}+\frac{(q-g_0(q))x}{r-r_s}=\Xi_1(x,r),
\label{theta1}
\end{eqnarray}
where 
\begin{eqnarray}
\Xi_1(x,r) &\equiv& \frac{\sum^N_{i=1}g_i(q)x^{\alpha_i}}{r-r_s}+h(q,x)(r-r_s)^a, \label{barxi}
\end{eqnarray}

We define the new coordinates as $u \equiv (r-r_s)^\beta$ and $x \equiv {\bar x}^\gamma$, where $\beta$ and $\gamma$ are positive constants.
Then Eq.~(\ref{theta1}) becomes
\begin{eqnarray}
\frac{d{\bar x}}{du}+\frac{q-g_0(q)}{\gamma\beta u}{\bar x}=\Xi_2({\bar x},u),
\label{theta2}
\end{eqnarray}
where $\Xi_2$ is defined by 
\begin{equation}
\Xi_2({\bar x},u) \equiv \frac{\sum^N_{i=1}g_i(q){\bar x}^{(\alpha_i-1)\gamma+1}}{\gamma \beta u}+\frac{h(q,x({\bar x}))}{\gamma \beta {\bar x}^{\gamma-1}}u^{(a-\beta+1)/\beta}.
\end{equation}
Thus, if $N=1$,  we choose $\gamma$ such that $\gamma=1/(1-\alpha_1)$ and then Eq.~(\ref{theta2}) is written as
\begin{eqnarray}
\frac{d{\bar x}}{du}+\frac{(1-\alpha_1)(q-g_0(q))}{\beta u}({\bar x}-\eta)=\eta \Xi_3({\bar x},u),
\label{theta3}
\end{eqnarray}
where we have introduced a parameter $0<\eta<\infty$ and $\Xi_3$ is defined by
\begin{eqnarray}
\Xi_3({\bar x},u) \equiv -\frac{(1-\alpha_1)[(q-g_0(q))\eta-g_1(q)]}{\eta\beta u}+\frac{(1-\alpha_1)h(q,x({\bar x}))}{\eta\beta {\bar x}^{\alpha_1/(1-\alpha_1)}}u^{(a-\beta+1)/\beta}.
\end{eqnarray}
Then, if we choose $\beta$ and $\eta$ such that $\beta \le (a+1)/2$ and $\eta= g_1(q)/(q-g_0(q)) \equiv \eta_0$, respectively, $\Xi_3$ is at least $C^1$ in $u \ge 0$ and ${\bar x}>0$.
Then we can apply the contraction mapping principle to Eq.~(\ref{theta3}) to find that there exists a solution satisfying ${\bar x}(0)=\eta_0$, and moreover that it is the unique solution of Eq.~(\ref{theta3}) which is continuous at $u=0$.
The proof is available in~\cite{christodoulou1984,newman1986,khi2000}.
$\Box$
\vspace*{1cm}

Thus, there is no other solution with $0<x(0)<\infty$ under the situations stated in Lemma~\ref{lm:fixedpoint}, so that other possible solutions must correspond to $x(0)=0$ or $\infty$. 
\begin{The}
Let us consider the minus-branch GB-LTB solution with positive mass and smooth initial profiles satisfying $\chi_2(0) > 0$.
Then, the singularity formed at $r=0$ is locally naked for $5 \le n \le 8$ and consequently the strong cosmic censorship hypothesis is violated.
\end{The}
{\it Proof.}
In order to find whether the singularities are naked or not, we investigate the geodesic equation for a future-directed outgoing radial null geodesic which emanates from the singularity.
We first adopt the so-called root-equation method to obtain the possible behavior of a null geodesic around $r=0$ if it exists~\cite{ns-dust,Joshi}; subsequently we show the existence of such a geodesic by Lemma~{\ref{lm:fixedpoint}}.

The constant $q$ in Eq.~(\ref{x}) must satisfy $q>1$ in this case, in which $r_s=0$, and is determined by requiring that $x$ has a positive finite limit $x_0$ for $r \to 0$ along the future-directed outgoing radial null geodesics.
It is noted that the regular center corresponds to $q=1$.
From the l'Hospital rule, we obtain
\begin{eqnarray}  
x_0 &=&\lim_{r \to 0}\frac{S}{r^{q}}\biggl|_{S=x_0r^q},\\
&=&\lim_{r \to 0}\left(\frac{1}{qr^{q-1}}\frac{dS}{dr}\biggl|_{+}\right)\biggl|_{S=x_0r^q},\\
&=&\lim_{r \to 0}\left(\frac{1}{qr^{q-1}}\left(S'+e^{\Psi-\Phi}{\dot S}\right)\right)\biggl|_{S=x_0r^q},\\
&=&\lim_{r \to 0}\left(\frac{v+rv'}{qr^{q-1}}\left(1+\frac{r{\dot v}}{\sqrt{1+r^2b}}\right)\right)\biggl|_{v=x_0r^{q-1}}.\label{x0}
\end{eqnarray}

Using Eqs.~(\ref{vdot}) and (\ref{v'}), we obtain from Eq.~(\ref{x0}) the desired root equation
\begin{eqnarray}  
x_0=\lim_{r \to 0}\frac{1}{qr^{q-1}}\left[v+\left(\frac{{\cal M}_0}{{\tilde\alpha}}\right)^{1/4}v^{(5-n)/4}r^2\chi_2(0)\right] \left[1-r\left(\frac{{\cal M}_0}{{\tilde\alpha}}\right)^{1/4}v^{(5-n)/4}\right]\biggl|_{v=x_0r^{q-1}} \label{root}
\end{eqnarray}
for $6 \le n \le 9$, while 
\begin{equation}  
x_0=\lim_{r \to 0}\frac{1}{qr^{q-1}}\left[v+\left(b_0+\sqrt{\frac{{\cal M}_0}{{\tilde\alpha}}}\right)^{1/2}r^2\chi_2(0)\right]\biggl|_{v=x_0r^{q-1}} \label{root50}
\end{equation}
for $n=5$.

For $n=5$, the positive finite root of Eq.~(\ref{root50}) is obtained with $q=3$ as
\begin{equation}  
x_0=\frac12\left(b_0+\sqrt{\frac{{\cal M}_0}{{\tilde\alpha}}}\right)^{1/2}\chi_2(0). \label{sol50}
\end{equation}
For $6 \le n \le 8$, it is obtained with $q=(n+7)/(n-1)>1$ as
\begin{equation}  
x_0=\left(\frac{n-1}{8}\right)^{4/(n-1)}\left(\frac{{\cal M}_0}{{\tilde\alpha}}\right)^{1/(n-1)}\chi_2(0)^{4/(n-1)}. \label{sol}
\end{equation}
For $n=9$, it is obtained with $q=2$ for 
\begin{eqnarray}  
\chi_2(0)\ge \frac{11+5\sqrt{5}}{2}\left(\frac{{\cal M}_0}{{\tilde\alpha}}\right)^{1/4} \label{condchi}
\end{eqnarray}
by solving the algebraic equation
\begin{eqnarray}  
x_0^3+\left(\frac{{\cal M}_0}{{\tilde\alpha}}\right)^{1/4}x_0^2-\chi_2(0)\left(\frac{{\cal M}_0}{{\tilde\alpha}}\right)^{1/4}x_0+\chi_2(0)\left(\frac{{\cal M}_0}{{\tilde\alpha}}\right)^{1/2}=0. \label{sol2}
\end{eqnarray}
Thus, we have found the possible behavior of a future-directed outgoing radial null geodesic which behaves near $r=0$ as
\begin{eqnarray}  
S \simeq x_0r^{(n+7)/(n-1)} \label{nullsol1}
\end{eqnarray}
for $5 \le n \le 9$, where $x_0$ is given by Eqs.~(\ref{sol50}), (\ref{sol}), and (\ref{sol2}) for $n=5$, $6 \le n \le 8$, and $n=9$, respectively.

We prove the existence of such a null geodesic by Lemma~\ref{lm:fixedpoint}. 
We consider the null geodesic equation (\ref{theta}) with $q=(n+7)/(n-1)$ for $5<n \le 9$.
By Eqs.~(\ref{vdot}) and (\ref{v'}), we expand $\Xi$ around $r=0$ as
\begin{eqnarray}
\Xi(x,r) &=& \frac{1}{r}\biggl[x+\left(\frac{{\cal M}_0}{\tilde\alpha}\right)^{1/4}x^{(5-n)/4}\chi_2(0)\biggl] -\left(\frac{{\cal M}_0}{\tilde\alpha}\right)^{1/4}x^{(5-n)/4}r^{2(5-n)/(n-1)}\biggl[x+\left(\frac{{\cal M}_0}{\tilde\alpha}\right)^{1/4}x^{(5-n)/4}\chi_2(0)\biggl] \nonumber \\
&&+\textrm{o}(r^{2(5-n)/(n-1)})
\end{eqnarray}
for $6 \le n \le 9$, while
\begin{eqnarray}
\Xi(x,r) = \frac{1}{r}\biggl[x+\left(b_0+\sqrt{\frac{{\cal M}_0}{{\tilde\alpha}}}\right)^{1/2}\chi_2(0)\biggl]+{\cal O}(r^{0}) 
\end{eqnarray}
for $n=5$.
We note that $2(5-n)/(n-1) \ge -1$ with equality holding for $n=9$.
Then, by Lemma~\ref{lm:fixedpoint}, we can show the existence of a null geodesic emanating from the singularity which behaves as Eq.(\ref{nullsol1}) around $r=0$, where $x_0$ is given by Eq.~(\ref{sol50}) and (\ref{sol}) for $n=5$, $6 \le n \le 8$, respectively. 
However, we cannot apply Lemma~\ref{lm:fixedpoint} to the case of $n=9$, so other methods are required to show the existence of the null geodesic in that case.

Finally we show that the curvature scalars actually diverge along the null geodesic, i.e., it is actually a singular null geodesic.
The energy density of a dust (\ref{GBLTB-rho}) is written as
\begin{eqnarray}  
\rho=\frac{[(n-1){\cal M}+r{\cal M}']}{V_{n-2}^1(v+rv')v^{n-2}}.\label{rhoex}
\end{eqnarray}
If $\rho$ diverges, the curvature scalars also diverge because Eq.~(\ref{beq}) gives
\begin{eqnarray}
\rho=\frac{1}{2\kappa_n^2}[(n-2)R+\alpha (n-4)L_{GB}]. \label{curvatures}
\end{eqnarray}
The shell-focusing singularity is here characterized by $v=0$, $r=0$, ${\cal M}={\cal M}_0$, and ${\cal M}'=0$ for $n \ge 5$, thus $\rho$ diverges if $rv'<\infty$ is satisfied for $r \to 0$ along the null geodesic (\ref{nullsol1}).
From Eq.~(\ref{v'}), we obtain
\begin{eqnarray}
v'\simeq \left(\frac{{\cal M}_0}{\tilde\alpha}\right)^{1/4}x_0^{(5-n)/4} r^{(9-n)/(n-1)}\chi_2(0)
\end{eqnarray}
for $6 \le n\le 9$, while
\begin{eqnarray}
v'\simeq \left(b_0+\sqrt{\frac{{\cal M}_0}{{\tilde\alpha}}}\right)^{1/2} r\chi_2(0)
\end{eqnarray}
for $n=5$ along the null geodesic (\ref{nullsol1}), so that we obtain $rv' \to 0$ for $r \to 0$.
Therefore, $\rho$ and the curvature scalars diverge along the null geodesics (\ref{nullsol1}).
$\Box$
\vspace*{1cm}

Unfortunately, the fixed-point method cannot be applied to show the existence of the singular null geodesics for $n=9$.
If they exist, they behave as Eq.~(\ref{nullsol1}) around $r=0$, where $x_0$ is given by Eq.~(\ref{sol2}). 

The general relativistic case has been investigated by several authors~\cite{gj2004}.
We review their analyses in Appendix~A for comparison with the case in Einstein-Gauss-Bonnet gravity and give some complements.

\subsubsection{Singularities at $r=r_s>0$ for $n=5$}
Next we consider singularities at $r=r_s$ with $0<r_s \le r_{\rm ah}$ in the five-dimensional case.
We expand ${\cal M}$ and $b$ around $r=r_s>0$ as
\begin{eqnarray}  
{\cal M}(r)&=&{\bar {\cal M}}_0+{\bar {\cal M}}_1(r-r_s)+{\bar {\cal M}}_2(r-r_s)^2+\cdots,\\
b(r)&=&{\bar b}_0+{\bar b}_1(r-r_s)+{\bar b}_2(r-r_s)^2+\cdots ,
\end{eqnarray}
where ${\bar {\cal M}}_0, {\bar {\cal M}}_1, {\bar {\cal M}}_2, \cdots$ and ${\bar b}_0, {\bar b}_1, {\bar b}_2, \cdots$ are constants.

Expanding $t(v,r)$ in Eq.~(\ref{t(vr)}) around $r=r_s$, we obtain
\begin{equation}  
t(v,r)=t(v,r_s)+(r-r_s){\bar \chi}_1(v)+\cdots, \label{texp5-2}
\end{equation}
where the function ${\bar \chi}_1(v)$ is defined by
\begin{equation}  
{\bar \chi}_1(v) \equiv -\int^1_v\frac{{\bar b}_1+{\bar {\cal M}}_1/\sqrt{v^{4}+4{\tilde\alpha}{\bar {\cal M}}_0}}{2\left[{\bar b}_0-\left(v^2-\sqrt{v^4+4{\tilde\alpha}{\bar {\cal M}}_0}\right)/(2{\tilde \alpha})\right]^{3/2}}dv. \label{chi25-2}
\end{equation}

The time $t={\bar t}_{s0}$ when the shell at $r=r_s$ reaches the singularity is given from Eq.~(\ref{t(vr)}) by
\begin{equation}  
{\bar t}_{s0}=\int^1_0\frac{dv}{\left[{\bar b}_0-\left(v^2-\sqrt{v^4+4{\tilde\alpha}{\bar{\cal M}}_0}\right)/(2{\tilde \alpha})\right]^{1/2}}. \label{ts0bar-2}
\end{equation}
From Eqs.~(\ref{texp5-2}) and (\ref{ts0bar-2}), the time $t=t_s(r)$ when other shells close to $r=r_s$ reach the singularity is given by
\begin{eqnarray}  
t_s(r) ={\bar t}_{s0}+(r-r_s){\bar \chi}_1(0)+\cdots. \label{ts-2}
\end{eqnarray}
By Lemma~\ref{lm:sing}, ${\bar \chi}_1(0)$ is non-negative and we assume ${\bar \chi}_1(0) > 0$ in this paper.
We show the nakedness of the singularity in the similar manner to the case of the singularity at $r=0$.
\begin{The}
In the five-dimensional minus-branch GB-LTB solution with positive mass and initial profiles satisfying ${\bar \chi}_1(0) > 0$, the singularity at $r=r_s$ with $0<r_s<r_{\rm ah}$ is locally naked and consequently the strong cosmic censorship hypothesis is violated.
\end{The}
{\it Proof.}
From Eq.~(\ref{x}) and the l'Hospital rule, we obtain
\begin{eqnarray}  
x_0 &=&\lim_{r \to r_s}\frac{S}{(r-r_s)^{q}}\biggl|_{S=x_0(r-r_s)^q},\\
&=&\lim_{r \to r_s}\left(\frac{1}{q(r-r_s)^{q-1}}\frac{dS}{dr}\biggl|_{+}\right)\biggl|_{S=x_0(r-r_s)^q},\\
&=&\lim_{r \to r_s}\left(\frac{v+rv'}{q(r-r_s)^{q-1}}\left(1+\frac{r{\dot v}}{\sqrt{1+r^2b}}\right)\right)\biggl|_{v=x_0(r-r_s)^{q}/r}.\label{x05}
\end{eqnarray}

From Eqs.~(\ref{t(vr)}), (\ref{dtdr}), and (\ref{texp5-2}), we obtain 
\begin{equation}  
v'=\left[{\bar b}_0-\frac{1}{2{\tilde \alpha}}\left(v^2-\sqrt{v^4+4{\tilde\alpha}{\bar{\cal M}}_0}\right)\right]^{1/2}({\bar\chi}_1(v)+\cdots) \label{v'5}
\end{equation}
near $r=r_s$.

From Eq.~(\ref{x05}) using Eqs.~(\ref{vdot}) and (\ref{v'5}), we obtain the desired root equation
\begin{eqnarray}  
x_0=\lim_{r \to r_s}\frac{r_s({\bar b}_0+\sqrt{{\bar {\cal M}}_0/{\tilde\alpha}})^{1/2}{\bar\chi}_1(0)}{q(r-r_s)^{q-1}}\left[1-\frac{({\bar b}_0+\sqrt{{\bar {\cal M}}_0/{\tilde\alpha}})^{1/2}}{({\bar b}_0+r_s^{-2})^{1/2}}\right].\label{root5-2}
\end{eqnarray}
The term in the large bracket in Eq.~(\ref{root5-2}) is non-negative because Eq.~(\ref{cr:cch0}) gives ${\bar b}_0+r_s^{-2} \ge {\bar b}_0+({\bar {\cal M}}_0/{\tilde\alpha})^{1/2}$ with equality holding for $r_s=r_{\rm ah}=({\bar {\cal M}}_0/{\tilde\alpha})^{1/2}$.
From Eq.~(\ref{root5-2}), we obtain 
\begin{equation}  
x_0=r_s{\bar\chi}_1(0)\left({\bar b}_0+\sqrt{{\bar {\cal M}}_0/{\tilde\alpha}}\right)^{1/2}\left[1-\frac{({\bar b}_0+\sqrt{{\bar {\cal M}}_0/{\tilde\alpha}})^{1/2}}{({\bar b}_0+r_s^{-2})^{1/2}}\right] \label{sol5-2}
\end{equation}
with $q=1$ for the singularity with $0<r_s <({\bar {\cal M}}_0/{\tilde\alpha})^{1/2}$.
Thus, we have found the possible behavior of a future-directed outgoing radial null geodesic which behaves near $r=r_s$ as
\begin{eqnarray}  
S \simeq x_0(r-r_s), \label{nullsol2}
\end{eqnarray}
where $x_0$ is given by (\ref{sol5-2}).

By Lemma~\ref{lm:fixedpoint}, we prove the existence of such a null geodesic. 
We consider the null geodesic equation (\ref{theta}) with $q=1$ for $n=5$.
By Eqs.~(\ref{vdot}) and (\ref{v'5}), we expand $\Xi$ around $r=r_s$ as
\begin{eqnarray}
\Xi(x,r) = \frac{r_s}{r-r_s}{\bar\chi}_1(0)\left({\bar b}_0+\sqrt{{\bar {\cal M}}_0/{\tilde\alpha}}\right)^{1/2}\left[1-\frac{({\bar b}_0+\sqrt{{\bar {\cal M}}_0/{\tilde\alpha}})^{1/2}}{({\bar b}_0+r_s^{-2})^{1/2}}\right]+{\cal O}((r-r_s)^{0}).
\end{eqnarray}
Then, by Lemma~\ref{lm:fixedpoint}, we can show the existence of a null geodesic which behaves as Eq.(\ref{nullsol2}) around $r=r_s$, where $x_0$ is given by Eq.~(\ref{sol5-2}).

Finally we show that it is actually a singular null geodesic.
Now the shell-focusing singularity is characterized by $v=0$, $r=r_s>0$, ${\cal M}={\bar {\cal M}}_0$, and ${\cal M}'={\bar {\cal M}}_1$.
From Eq.~(\ref{v'5}), we obtain
\begin{eqnarray}
v'&\simeq&\left({\bar b}_0+\sqrt{{\bar {\cal M}}_0/{\tilde\alpha}}\right)^{1/2}{\bar\chi}_1(0),
\end{eqnarray}
so that $rv'<\infty$ for $r \to r_s$ is satisfied along the null geodesic (\ref{nullsol2}).
As a result, we find from Eqs.~(\ref{rhoex}) and (\ref{curvatures}) that the curvature scalars diverge along this null geodesic (\ref{nullsol2}) and therefore it is a singular null geodesic.
$\Box$

\section{Discussion and conclusions}
In this paper, we have investigated the final fate of the gravitational collapse of a dust cloud in Einstein-Gauss-Bonnet gravity.
First, we have adopted the comoving coordinates and defined a scalar on $M^2$, of which dimension is mass, and given a simple formulation of the basic equations for the $n(\ge 5)$-dimensional spacetime $M \approx M^2 \times K^{n-2}$ including a perfect fluid and a cosmological constant in Einstein-Gauss-Bonnet gravity.
Then, having used this formalism in the spherically symmetric case without a cosmological constant, we have investigated the final fate of the $n(\ge 5)$-dimensional gravitational collapse of a dust cloud.
We have assumed that (i) the coupling constant of the Gauss-Bonnet term $\alpha$ is positive, which is required by heterotic superstring theory, (ii) positive mass of a dust fluid, (iii) smooth initial data around the symmetric center, and (iv) $S'>0$, i.e., no shell-crossing singularities. 

There are two families of solutions, which we call the plus-branch and the minus-branch GB-LTB solutions, respectively, although we have not obtained an explicit form of the solutions.
In the general relativistic limit ${\tilde \alpha} \to 0$, the minus-branch solution is reduced to the $n$-dimensional Lema{\^ i}tre-Tolman-Bondi solution~\cite{LTB,gj2004,higherdust}.
On the contrary there is no general relativistic limit for the plus-branch solution.
The GB-LTB solutions can be attached at the finite constant comoving radius $r=r_0>0$ to the outside vacuum spacetime represented by the GB-Schwarzschild solutions in the same branch with a mass parameter $M=m(r_0)$.

In the plus-branch solution with $n \ge 6$, bounce inevitably occurs and consequently singularities are never formed.
For $n=5$, a timelike naked singularity appears in the outside vacuum region, which is asymptotically anti-de Sitter in spite of the absence of a cosmological constant.
Since there is no trapped surface in the plus-branch solution, the singularity formed in the dust region must be naked, too. 

In the minus-branch solution, the outside vacuum region is asymptotically flat.
As in the case of a null dust fluid~\cite{maeda2005}, the final fate of gravitational collapse is quite different depending on whether $n=5$ or $n \ge 6$.
A massive naked singularity is formed for $n=5$, which is prohibited for $n \ge 6$ and in general relativity. 

In the outside vacuum region, a massive timelike naked singularity appears if the mass of a dust cloud satisfies $0<m(r_0)<3{\tilde\alpha}V_3^1/(2\kappa^2_5)$ for $n=5$.
If either $n \ge 6$ or $n=5$ with $m(r_0)>3{\tilde\alpha}V_3^1/(2\kappa^2_5)$ is satisfied, the singularity is spacelike.
In the special case of $n=5$ with $m(r_0)=3{\tilde\alpha}V_3^1/(2\kappa^2_5)$, it is outgoing-null.

We have treated the homogeneous and inhomogeneous cases in the minus-branch solution, separately.
In the homogeneous collapse represented by the $n$-dimensional flat Friedmann-Robertson-Walker solution, the singularity formed in this Friedmann region is censored, which is spacelike for $n \ge 6$ and ingoing null for $n=5$.
As a result, the strong CCH holds for $n \ge 6$ or $n=5$ with $m(r_0) \ge 3{\tilde\alpha}V_3^1/(2\kappa^2_5)$, while the weak CCH is violated for $n=5$ with $0<m(r_0)<3{\tilde\alpha}V_3^1/(2\kappa^2_5)$.
The global structure for $n \ge 6$ is the same as that of the general relativistic case, which is the $n$-dimensional Oppenheimer-Snyder solution representing black-hole formation.

In the inhomogeneous collapse with smooth initial data, the strong CCH holds for $n \ge 10$.
For $n=9$, the strong CCH holds depending on the parameters in the initial data.
On the other hand, a naked singularity is formed in the dust region for $5 \le n \le 8$.
The naked singularity is massless for $6 \le n \le 8$, while it is massive for $n=5$.
Thus, at least the strong CCH is violated for $6 \le n \le 8$ or $n=5$ with $m(r_0) \ge 3{\tilde\alpha}V_3^1/(2\kappa^2_5)$, while the weak CCH is violated for $n=5$ with $0<m(r_0)<3{\tilde\alpha}V_3^1/(2\kappa^2_5)$.
Although we have not solved the null geodesic equation in the whole spacetime, the singularity can be globally naked if we take the limit $r_0 \to 0+\varepsilon$ for $6 \le n \le 8$ and $r_0 \to r_{\rm ah}+\varepsilon$ for $n=5$, where $\varepsilon$ is a sufficiently small positive constant such that a light ray emanating from the singularity reaches the surface $r=r_0$ in the untrapped region.
Then the light ray can escape to infinity and consequently the weak CCH is violated.

Unfortunately, we could not show the existence of singular null geodesics for $n=9$ although we could show their possible behavior near the singularity.
The situation is the same for $n=5$ in general relativity, as shown in Appendix A.
Other approaches such as the comparison method are required to show the nakedness of the singularity in these cases~\cite{nm2002}.

In general relativity~\cite{LTB,gj2004} with smooth initial data, the weak CCH holds for $n \ge 6$~\cite{LTB,gj2004}.
The present result and the result in~\cite{maeda2005} imply that the effects of the Gauss-Bonnet term worsen the situation from the viewpoint of CCH rather than prevent naked singularity formation.

The formation of massive naked singularities in Einstein-Gauss-Bonnet gravity for $n=5$ is quite remarkable.
In four dimensions with spherical symmetry, it has been shown by Lake under very generic situations without using the Einstein equations that massive singularities formed from regular initial data are censored by adopting the Misner-Sharp mass as a quasi-local mass~\cite{lake1992}.
As was already pointed out in~\cite{maeda2005}, the higher-dimensional counterpart of the Misner-Sharp mass is not an appropriate quasi-local mass in Einstein-Gauss-Bonnet gravity, and therefore his result does not conflict with ours. 
Lemma~\ref{lm:mass} implies that, in Einstein-Gauss-Bonnet gravity, our definition of the quasi-local mass is preferable to the higher-dimensional Misner-Sharp mass.
Lake's result must be extended in Einstein-Gauss-Bonnet gravity for $n \ge 6$ by adopting our quasi-local mass.

The formation of massive singularities in odd dimensions is considered to be a characteristic effect of Lovelock terms in the action, which include the Gauss-Bonnet term as the quadratic term~\cite{nm2005}.
The Gauss-Bonnet term becomes first nontrivial in five dimensions, so that massive singularities can be formed only in five dimensions.
If we add the higher-order Lovelock terms in the action, massive naked singularities must be formed in all odd dimensions.

In this paper, we have assumed that the equation of state of matter is dust in the final stages of gravitational collapse.
Although it is a very strong assumption, it is not completely ruled out~\cite{Hagadorn}.
Actually, we have little knowledge of the equation of state in the very advanced stage of gravitational collapse. 
However, of course, the effects of pressure should be investigated.
In general relativity, there have been many studies including pressure both in four dimensions~\cite{ns-pf4,ns-scalar} and in higher dimensions~\cite{higherpressurefluid,highernumerical,highertypeII,higherscalar}. 

Among them, a scalar field must be especially addressed in the higher-dimensional context because it arises naturally in supergravity~\cite{sugra} and plays a central role in modern cosmology~\cite{linde}.
In the brane-world scenario, a scalar field may exist in the bulk spacetime.
In four dimensions, the CCH has been proven in spherically symmetric collapse of a massless scalar field in general relativity~\cite{ns-scalar}.
This result will hold in higher dimensions because higher-dimensional effects make singularities censored.
However, it's just conceivable that the effects of the Gauss-Bonnet term will change drastically the whole picture and the final fate of gravitational collapse.

Among scalar fields, a dilaton field is particularly of interest because it naturally arises in the low-energy effective theory of superstring theory~\cite{Gross,superstring}.
Indeed, it is coupled to the Gauss-Bonnet term, and consequently the Gauss-Bonnet term does contribute to the field equations even in four dimensions~\cite{Gross,GBD}.
In any case, the quasi-local mass defined in this paper should be useful in the study of gravitational collapse in Einstein-Gauss-Bonnet gravity.

Lastly, several notes should be made.
In this paper, we have assumed smooth initial data around the symmetric center.
Analyses with more general initial data are needed to determine whether the nature of the singularities and the picture of gravitational collapse obtained in this paper are generic or not.
Also, the structure and the strength of the naked singularities are still open.
In the case of a null dust, the Gauss-Bonnet term weakens the strength of the naked singularity and the naked singularity is timelike and ingoing null for $n=5$ and $n \ge 6$, respectively.
These studies will be reported elsewhere.

{\it Acknowledgements:}
The author would like to thank M.~Narita, T.~Harada, and M.~Nozawa for discussions and useful comments.

\appendix
\section{general relativistic case}
In this appendix, we review the general relativistic case for $n \ge 4$ and give some complements to the previous studies for $n \ge 5$~\cite{gj2004}.
The equations in general relativity can be obtained from those in Einstein-Gauss-Bonnet gravity by the limit of ${\tilde\alpha} \to 0$.

First we consider the distribution of the apparent horizon around $r=0$.
From Eq.~(\ref{trap}), the trapped region is given by
\begin{eqnarray}
{\cal M}>r^{-2}v^{n-3}. \label{trap2-gr}
\end{eqnarray} 
From Eq.~(\ref{ah}), the apparent horizon is given by
\begin{eqnarray}
{\cal M}=r^{-2}v^{n-3}, \label{ah2-gr}
\end{eqnarray} 
so that only the singularity at $r=0$ may be naked, in other words, no massive naked singularity exists.

We derive the relation between the time when the singularity occurs and the time when the apparent horizon appears for each $r$.
From Eq.~(\ref{vdot}), we obtain
\begin{eqnarray}  
{\dot v}=-\left(b+{\cal M}v^{3-n}\right)^{1/2}. \label{vdot-gr}
\end{eqnarray}
Integrating this equation with respect to $v$, we obtain
\begin{equation}  
t(v,r)=\int^1_v\frac{dv}{\left(b+{\cal M}v^{3-n}\right)^{1/2}}, \label{t(vr)-gr}
\end{equation}
where we have used the freedom of the radial coordinate $r$ and of the time translation so that $S=r$ is satisfied at the initial moment $t=0$.

The time $t=t_s(r)$ when each shell reaches the singularity is given from Eq.~(\ref{t(vr)-gr}) by
\begin{equation}  
t_s(r) = \int^1_0\frac{dv}{\left(b+{\cal M}v^{3-n}\right)^{1/2}}. \label{ts-zero-gr}
\end{equation}
From Eq.~(\ref{t(vr)-gr}), we also obtain the time $t=t_{\rm ah}(r)$ when the apparent horizon appears for each $r$ by
\begin{eqnarray}  
t_{\rm ah}(r)&\equiv&\int^1_{v_{\rm ah}(r)}\frac{dv}{\left(b+{\cal M}v^{3-n}\right)^{1/2}}, \nonumber \\
&=&\int^1_0\frac{dv}{\left(b+{\cal M}v^{3-n}\right)^{1/2}}+\int^0_{v_{\rm ah}(r)}\frac{dv}{\left(b+{\cal M}v^{3-n}\right)^{1/2}}, \nonumber \\
&=&t_{s}(r)-\int^{v_{\rm ah}(r)}_0\frac{dv}{\left(b+{\cal M}v^{3-n}\right)^{1/2}}, \label{tah1-gr}
\end{eqnarray}
where $v=v_{\rm ah}(r)$ is given by solving Eq.~(\ref{ah2-gr}).

Expanding Eq.~(\ref{t(vr)-gr}) around $r=0$, we obtain
\begin{equation}  
t(v,r)=t(v,0)+\frac{r^2}{2}\chi_2(v)+\cdots, \label{texp-gr}
\end{equation}
where the function $\chi_2(v)$ is now obtained by
\begin{equation}  
\chi_2(v) \equiv -\int^1_v\left(b_0+{\cal M}_0v^{3-n}\right)^{-3/2}\left(b_2+{\cal M}_2v^{3-n}\right)dv,
\end{equation}
where the integrand is finite for $0 \le v \le 1$.
The time when the central shell reaches the singularity is given from Eq.~(\ref{t(vr)-gr}) by
\begin{equation}  
t_{s0}=\int^1_0\frac{dv}{\left(b_0+{\cal M}_0v^{3-n}\right)^{1/2}}. \label{ts0-gr}
\end{equation}
From Eqs.~(\ref{texp-gr}) and (\ref{ts0-gr}), the time when other shells close to $r=0$ reach the singularity is given by
\begin{eqnarray}  
t_s(r) =t_{s0}+\frac{r^2}{2}\chi_2(0)+\cdots. \label{ts-gr}
\end{eqnarray}
By Lemma~\ref{lm:sing}, $\chi_2(0)$ is non-negative and we assume $\chi_2(0) > 0$ here.

From Eqs.~(\ref{t(vr)-gr}), (\ref{texp-gr}) and (\ref{dtdr}), we obtain 
\begin{eqnarray}  
v'=\left(b_0+{\cal M}_0v^{3-n}\right)^{1/2}(r\chi_2(v)+\cdots) \label{v'-gr} 
\end{eqnarray}
near $r=0$.
\begin{The}
Let us consider the gravitational collapse of a spherical dust cloud with positive mass and smooth initial profiles satisfying $\chi_2(0) > 0$ in general relativity without a cosmological constant.
Then, the strong cosmic censorship hypothesis holds for $n \ge 6$ and $n=5$ if $0<\chi_2(0)<{\cal M}_0^{1/2}$. 
\end{The}
{\it Proof.}
From Eq.~(\ref{trap2-gr}), only the singularity at $r=0$ has the possibility of being naked.  
A decreasing apparent horizon in the $(r,t)$-plane is a sufficient condition for the formed singularity to be censored because it shows the entrapment of the neighborhood of the center before the singularity.
Expanding the integrand in Eq.~(\ref{tah1-gr}) in a power series in $v$ and keeping only the leading order term, we obtain
\begin{eqnarray}  
t_{\rm ah}(r)&=&t_{s0}+\frac{r^2}{2}\chi_2(0)+\cdots  \nonumber \\
&&-\frac{2}{n-1}{\cal M}_0^{1/(n-3)}r^{(n-1)/(n-3)}.
\end{eqnarray}
For $n \ge 6$, the last term dominates the second term, and consequently the apparent horizon is decreasing near $r=0$ in the $(r,t)$-plane.
For $n=5$ with $\chi_2(0)<{\cal M}_0^{1/2}$, the apparent horizon is also decreasing near $r=0$ in the $(r,t)$-plane.
$\Box$
\vspace*{1cm}

Next we consider the nakedness of singularities for $n=4$ and $5$.
We show the nakedness of singularities in the similar manner to the case in Einstein-Gauss-Bonnet gravity.
\begin{The}
Let us consider the $n(\ge 4)$-dimensional gravitational collapse of a spherical dust cloud with positive mass and smooth initial profiles satisfying $\chi_2(0) > 0$ in general relativity without a cosmological constant.
Then, the strong cosmic censorship hypothesis is violated for $n=4$.
\end{The}
{\it Proof.}
Using Eqs.~(\ref{vdot-gr}) and (\ref{v'-gr}), we obtain from Eq.~(\ref{x0}) the desired root equation
\begin{eqnarray}  
x_0=\lim_{r \to 0}\frac{1}{qr^{q-1}}\left(v+{\cal M}_0^{1/2}v^{(3-n)/2}r^2\chi_2(0)\right)\left(1-r{\cal M}_0^{1/2}v^{(3-n)/2}\right)\biggl|_{v=x_0r^{q-1}}. \label{root-gr}
\end{eqnarray}

For $n=4$, the positive finite root of Eq.~(\ref{root-gr}) is obtained with $q=(n+3)/(n-1)=7/3$ as
\begin{equation}  
x_0=\left(\frac{3{\cal M}_0^{1/2}\chi_2(0)}{4}\right)^{2/3}. \label{sol4-gr}
\end{equation}
For $n=5$, it is obtained with $q=(n+3)/(n-1)=2$ for 
\begin{eqnarray}  
\chi_2(0)\ge \frac{11+5\sqrt{5}}{2}{\cal M}_0^{1/2} \label{condchi-gr}
\end{eqnarray}
by solving the algebraic equation
\begin{equation}  
x_0^3+{\cal M}_0^{1/2}x_0^2-\chi_2(0){\cal M}_0^{1/2}x_0+\chi_2(0){\cal M}_0=0. \label{sol5-gr}
\end{equation}
Thus, we have found the possible behaviors of a future-directed outgoing radial null geodesic near $r=0$ as
\begin{eqnarray}  
S \simeq x_0r^{(n+3)/(n-1)}, \label{nullsol1-gr}
\end{eqnarray}
where $x_0$ is given by Eqs.~(\ref{sol4-gr}) and (\ref{sol5-gr}) for $n=4$ and $n=5$, respectively.

By Lemma~\ref{lm:fixedpoint}, we prove the existence of such a null geodesic. 
We consider the null geodesic equation (\ref{theta}) with $q=(n+3)/(n-1)$.
By Eqs.~(\ref{vdot-gr}) and (\ref{v'-gr}), we expand $\Xi$ around $r=0$ as
\begin{eqnarray}
\Xi(x,r) &=& \frac{1}{r}\biggl[x+{\cal M}_0^{1/2}x^{(3-n)/2}\chi_2(0)\biggl]-{\cal M}_0^{1/2}x^{(3-n)/2}r^{2(3-n)/(n-1)}\biggl(x+{\cal M}_0^{1/2}x^{(3-n)/2}\chi_2(0)\biggl) \nonumber \\
&&+\textrm{o}(r^{2(3-n)/(n-1)}).
\end{eqnarray}
We note that $2(3-n)/(n-1) \ge -1$ with equality holding for $n=5$.
Then, by Lemma~\ref{lm:fixedpoint}, we can show the existence of a null geodesic which behaves as Eq.~(\ref{nullsol1-gr}) around $r=0$ for $n=4$, where $x_0$ is given by Eq.~(\ref{sol4-gr}). 
However, as the case of $n=9$ in Einstein-Gauss-Bonnet gravity, we cannot apply Lemma~\ref{lm:fixedpoint} to the case of $n=5$ in general relativity, so another method is required to show the existence in that case.

Finally we show that the curvature scalars actually diverge along the null geodesic.
If $\rho$ diverges, the curvature scalars also diverge because Eq.~(\ref{beq}) gives
\begin{eqnarray}
\rho=\frac{(n-2)R}{2\kappa_n^2} \label{curvatures-gr}
\end{eqnarray}
in general relativity. 
The shell-focusing singularity is characterized by $v=0$, $r=0$, ${\cal M}={\cal M}_0$, and ${\cal M}'=0$ for $n \ge 4$.
From Eq.~(\ref{v'-gr}), we obtain
\begin{eqnarray}
v'\simeq {\cal M}_0^{1/2}x_0^{(3-n)/2} r^{(5-n)/(n-1)}\chi_2(0)
\end{eqnarray}
along the null geodesic (\ref{nullsol1-gr}), so that $rv' \to 0$ for $r \to 0$.
Therefore, $\rho$ and the curvature scalars diverge along the null geodesic (\ref{nullsol1-gr}).
$\Box$
\vspace*{1cm}

Unfortunately, the fixed-point method cannot be applied to show the existence of the singular null geodesics for $n=5$.
If they exist, they behave as Eq.~(\ref{nullsol1-gr}) around $r=0$, where $x_0$ is given by Eq.~(\ref{sol5-gr}).


\end{document}